# "It takes two to tango".
# A Review of the Empirical Literature on Information Technology Outsourcing Relationship Satisfaction.

Draft paper[1]
Ph.D. dissertation
Jorg Verbaas

Tilburg University

Date: 7 November 2010
Version: 0.6

**PhD student:**
Name: Jorg Verbaas
E-mail: jorg.verbaas@accenture.com

---

[1] This paper is part of a PhD project that was sponsored by Logica NL (drs. E.H.S. Mensonides) and has been written in association with Prof. dr. L.A.G. Oerlemans and dr. J.M.J. Baaijens. The author is grateful for their research support.

# "It takes two to tango". A Review of the Empirical Literature on Information Technology Outsourcing Relationship Satisfaction

Jorg Verbaas

**Abstract**

There is growing recognition that the overall client-vendor relationship, and not only the contract, plays a critical role in Information Technology Outsourcing (ITO) success. However, our understanding of how ITO relationships function is limited. This paper contributes to this understanding by reviewing empirical literature on ITO success in terms of relationship satisfaction. A key finding is that the majority of reviewed studies concentrates on *client satisfaction*, thus neglecting the vendor perspective. We argue that this raises questions about the *construct validity* of these studies. Consequently, concerns exist about the validity and reliability of their empirical findings. Some scholars have acknowledged the problem and use a *dyadic perspective*. However, a review of these studies reveals that the authors have underestimated their contributions and do not explain *why* there is a problem. Therefore, the purpose of this paper is to highlight their contributions by comparing the findings of the dyadic perspective studies with those of the "client perspective" research. In doing so, we assess whether the dyadic studies produce better explanations for ITO success than the client-oriented studies. We argue that this is indeed the case, by producing a better view on how underlying mechanisms of ITO relationships work.

## Introduction

Information Technology Outsourcing (ITO) has become a generally accepted option for client organisations to satisfy their organisational information systems (IS) needs (Goles, 2003: 199) and is an almost omnipresent phenomenon by now. Reported worldwide ITO revenues in 2005 vary from $152 billion (Lee et al., 2008) to over $200 billion (Seddon et al., 2007). However, ITO is not without failure. It is reported that 34% of outsourcing is brought back in-house (Lacity and Willcocks, 2000 in: Whitten, 2004: 2). A Deloitte study (Landis et al., 2005) found that 70 percent of clients have had significant negative experiences with their outsourcing arrangement; 64% of the researched organisations brought outsourced services back in-house. Bravard and Morgan (2006) indicate that roughly two-thirds of outsourcing agreements fail to deliver the targeted benefits.

There is growing recognition that the overall relationship between a client and vendor, and not only the contractual terms, plays a critical role in the success or failure of ITO (e.g., Goles and Chin, 2005; Kern and Willcocks, 2000; Jahner et al., 2006; Sargent, 2006). This is in part because clients increasingly shift focus from cost savings to outsourcing as enabler for improving their overall business performance and competitive position (cf. Goles and Chin, 2005: 47), which may increase the strategic value of the relationship with a vendor. Moreover, clients increasingly realize that well-designed contracts alone are insufficient to successfully manage vendors. An increasing number of clients therefore tends to seek more partnership-based relationships with their service providers based on mutual trust (cf. Lee and Kim, 2005). However, our understanding of how ITO relationships function is limited (Kern and Willcocks, 2002; Goles and Chin, 2005). Hence, we need to learn more about these relationships.

Much of the ITO literature concentrates on *outcomes* of ITO arrangements such as stated above, in which most scholars have focused on the factors that impact the *success* of outsourcing (Dibbern et al., 2004: 69). However, these factors are heavily dependent on how one defines "outcome" or "success". A commonly used proxy for indicating overall outsourcing success is *satisfaction* (e.g., Goles, 2003: 202; Koh et al., 2004: 366). Yet, the majority of studies concerned with examining factors that impact ITO success in terms of satisfaction concentrates on *client satisfaction* (see table 5a in Appendix B). This means that these studies focus exclusively on one side of an ITO relationship, thus discarding the vendor's perspective.

Although not all types of ITO relationships may be coined "strategic partnerships" (cf. Kern, 1997: 48), and thus may be seen as arm's length relationships, at their most basic level all ITO arrangements involve at least two *participants* in some type of *exchange relationship* (Goles, 2001: 30; Lee et al, 2008: 147). Therefore, in order to study ITO relationships it is pivotal to examine the exchange relationship – e.g., interactions and transactions – between the two participants (the client and vendor). This *relational perspective* is based on the recognition of mutuality of the parties involved in an ITO relationship. Mutuality is anchored in the reciprocal relationship between the two parties, in which the supplier agrees to make specific contributions to the client in return for certain benefits – e.g., payment – from the client (Koh et al., 2004: 357).

As will be argued, the majority of the studies regarding overall satisfaction with ITO arrangements have one-sidedly focused on the clients' perspective. To counterbalance this unilateral focus on clients, a number of scholars have examined ITO from the vendor's perspective only (e.g., Levina and Ross, 2003; Oza et al., 2006). So, helpful as it may seem, the latter approach has in common with the unilateral focus on clients that it views ITO relationships from one side of the equation only.

Some have compared ITO to marriage (e.g., Gong et al, 2007; Lee and Kim, 2005: 44). Following this analogy, the above unilateral perspectives look somewhat like evaluating the success of a marriage through interviewing the husband or wife only. In other words, we believe that the majority of prior research on ITO success has failed to recognize or at least has not acted upon the fact that in order to assess the success of an ITO arrangement one has to examine both clients' and vendors' perspectives, i.e. take a *dyadic perspective*.

We argue that most research on ITO success raises questions about the *construct validity* – defined broadly as the extent to which an operationalisation of a construct measures the concept it is supposed to measure (e.g., Bagozzi et al., 1991: 421; John and Reve, 1982: 520) – of the literature regarding satisfaction with ITO relationships. As a result, there are concerns about the validity and reliability of the empirical findings of the studies at hand.

A small number of scholars (see table 5b in Appendix B) has acknowledged the problem and uses a dyadic perspective. However, a review of these studies reveals that the authors themselves have underestimated their contributions. Therefore, the main purpose of this paper is to highlight their contributions by comparing the findings of the small number of "dyadic perspective" studies with those of the larger number of "client perspective" research. In doing so, we assess whether the dyadic studies produce better and/or other explanations for ITO success than the client-oriented studies. To this end we have developed evaluation criteria with which to carry out this assessment. Consequently, the research question for this paper is: *do studies with a dyadic perspective on ITO relationships lead to better and/or other explanations for ITO success than studies that focus on the client perspective only?* In addition, this paper provides a state-of-the-art overview of studies on satisfaction with ITO relationships, thus answering to the call for review articles in the IS field and contributing to a synthesis of prior research (cf. Webster and Watson, 2002; Dibbern et al., 2004).

The remainder of this paper is structured as follows. First, we explain our view of a relational perspective and its benefits. Next, we will introduce the satisfaction concept and will present a definition for this concept that has been used as a lens to review the existing literature. The third section explains the research method used in this study. The Results section explores in detail what we have found from the reviewed literature. The final two sections offer recommendations for future research and cover the contributions of this paper and its limitations respectively.

## Relational perspective

This study applies a *relational perspective*. This section discusses what it is, and why it is important to use a relational perspective.

### *What is a relational perspective?*

ITO is one form of inter-organisational relationships (IOR). According to Oliver (1990: 241) IORs are "the relatively enduring transactions, flows and linkages that occur among or between an organisation and one or more organisations in its environment". From this definition we may learn that IORs can be observed from at least five points of view: a) the focal organisation – the ego; b) its partner(s) – the alter(s); c) their exchange relationship; d) the context or environment; and e) the time dimension.

These five points of view constitute the building blocks of the relational perspective that we use in this paper. In this perspective relationships are not studied "atomistically" from the focal organisation's perspective and its attributes, such as firm size and firm-specific assets (Poppo and Zenger, 1998). Rather, the relational perspective views focal organisations – predominantly the clients in ITO literature – as embedded in and consisting of internal and external networks of relationships (cf. Brass et al., 1998: 17) with employees, suppliers, institutional actors such as governments, regulatory bodies, professional associations, etc. They therefore are not studied in isolation. In the relational perspective, it is believed that relationships and their characteristics (e.g., the level of exchanges, trust or knowledge transfer) are relevant for understanding organisational behaviour and outcomes. For example, relationships co-determine survival chances of organisations because the relationships enable access to complementary resources, with which to develop adaptive strategies to cope with competitive and institutional pressures.

Second, the relational perspective offers a different view on environments. It translates abstract and faceless environments into agents that can take on different roles like (pro)active stakeholders and shareholders (cf. Nohria (1992: 5). It is this reframing of environments that allows for an interaction, exchange and agency perspective on the environment. This is important because the environment of organisations does not merely consist of abstract trends monitored cognitively and translated into strategies; it consists of competitors, suppliers, and customers, regulatory agencies, and other institutional actors with whom organisations can or cannot interact in order to respond to environmental pressures.

The relational perspective includes the time dimension in at least three respects. First, ITO relationships do not end after a single transaction but under certain conditions will continue over a longer period of time. Thus, they involve a process, more specifically "an ongoing reciprocal process, in which actions from one ITO partner are contingent on rewarding reactions from the other partner" (cf. Blau, 1964 in: Das and Teng, 2002: 448). This refers to the mutuality of the parties involved in an ITO relationship. We will discuss this element in the next subsection. Second, during this ongoing reciprocal process, the perception of the relationship by both partners, e.g., in terms of satisfaction, may fluctuate over time. Without a longitudinal approach, this aspect of a relationship cannot be captured. Third, the parties involved may not always be in agreement about how the relationship develops. In case of a single transaction, this problem may be resolved quickly, either by a satisfactory solution to both parties or by a termination of the transaction. In case of an ITO relationship this is more difficult, due to investments made in the relationship, high switching costs, etc.

For instance, ITO partners may both be satisfied at the beginning of their relationship: the vendor has won a contract with a high profile client; the client is pleased with the prospect of significant cost savings. Levinthal and Fichman (1988) have coined this the "honeymoon" period. Then, let us assume that after a few months the vendor is unable to meet the agreed-upon requirements and that the client becomes dissatisfied. The vendor may feel unhappy about that and may respond by deploying more personnel, changing key staff, etc., and the quality of delivered services increases. Or, the vendor may disagree and do nothing at all. In turn, the client will respond to the vendor's reaction and may become satisfied again or even more dissatisfied, etc.

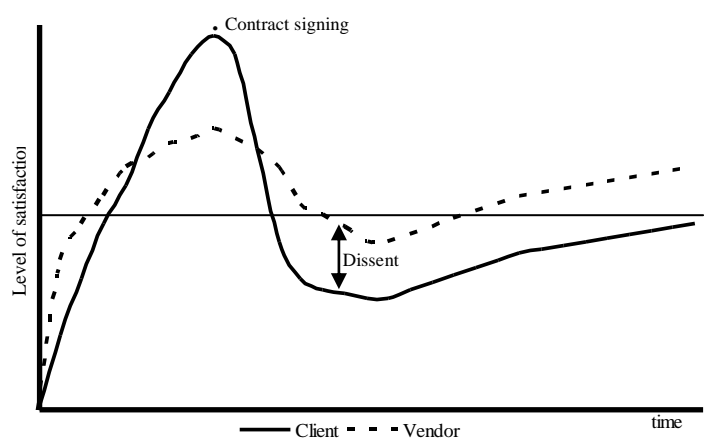

**Figure 1** Levels of relationship satisfaction, as perceived by clients and vendors, over time

Note: This is a purely theoretical illustration. The thin horizontal line represents an average level of satisfaction.

These fluctuations in levels of satisfaction will probably continue throughout the lifespan of the relationship. See figure 1 for an illustration. In addition to these fluctuations *per se*, a relational perspective allows for the possibility that the partners may perceive the satisfaction with their ITO relationship differently at various points in time. See the differences in the two curves in figure 1. In our example, this difference in perception may occur at the point where the vendor disagrees with the client. In case the client would not express its dissatisfaction, and so the vendor may be unaware of this fact, the vendor might think that the ITO arrangement is going well. We coin this kind of difference of opinion *dissent*, which already has been examined in settings similar to ITO. In the context of IT projects, Meeus and Oerlemans (2002: 190) have defined "level of dissent" as "the extent to which stakeholders agree on intervention outcomes". In their study, the stakeholders were managers and employees respectively, who may judge outcomes differently. In our case the stakeholders are clients and vendors, and the intervention outcome in the above example is satisfaction.

Consistent with the above discussed relational perspective, and for the purpose of this paper, we will use Goles and Chin's (2005: 49) definition of an ITO relationship: *an ongoing, long term linkage between an outsourcing vendor and customer arising from a contractual agreement to provide one or more comprehensive IT activities, processes, or services with the understanding that the benefits attained by each firm are at least in part dependent on the other*. For the purpose of this paper we only consider *dyadic* ITO relationships. We are aware that this is a theoretical viewpoint, as ITO relationships frequently consist of more than two partners – e.g., see Tikkanen et al. (2000: 380) for a "network approach" towards ITO relationships. The unit of analysis in this study is thus an ITO relationship between a client and vendor, viewed from both the service receiver's and provider's perspectives.

## *Why is it important to use a relational perspective?*

We contend that when studying ITO relationships, and for instance analysing the variables that influence their success, it is imperative to use a relational perspective. This is illustrated by the inherent mutual nature of relationship variables. Without a relational perspective this mutual nature cannot be captured. Relationship variables are variables that cease to exist if a relationship were terminated (Oerlemans et al., 2007: 206). They are frequently used to explain the success of ITO relationships.

For instance, let us examine the role of trust in ITO relationships. Lee et al. (2008) have found statistical evidence that initial trust and distrust by both sides of an ITO relationship influence mutual trust at a later stage. It may therefore indicate that the level of trust of one participant influences the other participant's level of trust. Ring and Van de Ven (1994) and Doz (1996) argue that this is due to a cyclical process IORs undergo, in which renegotiation or re-evaluation leads to adjustments in mutual expectations, trust and commitments.

Identically, communication is a mutual, two-way process (Goles and Chin, 2005: 56), i.e. communication behaviour from one participant influences communication behaviour of the other. For example, if a client expresses negative emotions in its communication, the vendor may respond in an equally negative way (cf. Celuch et al, 2006: 575).

The interdependence concept is another example. According to Anderson and Narus (1990: 43) this concept explains a firm's dependence *relative to* its partner's dependence on an IOR. That is, it takes into account the balance of power in an IOR. For example, if firm A is less dependent on the relationship than firm B, firm A can use this position to bring about better outcomes for itself. In other words, interdependence is also a mutual concept.

As may be derived from these three examples, ITO relationships cannot be examined adequately from one point of view – e.g., the focal organisation – alone. In that case relevant information would be neglected, such as the mutual effect of communication and trust. Moreover, these effects take place *over time*, which means that the study of ITO relationships would benefit from using a longitudinal approach, as discussed earlier.

## Introducing the satisfaction concept

As mentioned, a commonly used proxy for indicating overall outsourcing success is *satisfaction*. We will introduce this concept here, since this literature review concentrates on research that has used it as (a proxy for) the dependent variable. Furthermore, the satisfaction definition that we have developed below has been used to determine the scope of the reviewed studies.

Satisfaction is a commonly used construct, especially in the (consumer) marketing literature. It has been found to have a strong link with repurchase intentions and customer retention (cf. Susarla et al., 2003: 93; Koh et al., 2004: 366), making it an important measure for success. Given the increasing number of articles on ITO that incorporates the satisfaction concept, the construct appears to have gained a strong foothold in the ITO literature as well. Satisfaction is indicated as an effective overall measure because it allows ITO arrangements to be assessed through criteria that are the most relevant to the participants (Goles, 2003: 202; Seddon et al., 2007: 243). For instance, the measurement "are you satisfied with your ITO arrangement?" gives a respondent the freedom to fill in its own organisation's goals, expectations, etc. rather than having to respond to a predefined set of criteria. From a methodological point of view, the downside of this approach would be that it is unclear what specific object of satisfaction is being researched. In order to overcome this problem a multi-faced construct is needed. We will return to this point in the Results section, where we address the issue of defining and measuring the "ITO success" concept.

Several scholars view "satisfaction with one's partner" in dyadic relationships as an adequate indicator or even a substitute measure for success. For instance, some suggest that a partner's satisfaction with the other across several aspects of the relationship is an indicator of partnership success (Mohr and Spekman, 1994: 142) and a "close proxy" for perceived effectiveness of partnerships (Anderson and Narus, 1990: 46). Others (e.g., Lee and Kim, 1999; Grover et al., 1996; Saunders et al., 1997) have measured ITO success in terms of satisfaction. Dibbern et al. (2004: 74) and Goles (2001: 77) contend that satisfaction is a "surrogate" for ITO success. Given the importance of the satisfaction concept in the ITO literature, we will explore this concept in detail below. We will, however, position satisfaction in the context of ITO relationships, thus focusing on *ITO relationship satisfaction*.

Relationship satisfaction has been defined as "a global evaluation of fulfilment in the relation" (Dwyer and Oh, 1987: 352) and as "a positive affective state based on the outcomes obtained from the relationship" (Ganesan, 1994: 4). It encompasses all characteristics of a relationship that a focal firm finds rewarding, profitable or instrumental on the one hand or frustrating, problematic, or inhibiting on the other (Ruekert and Churchill, 1984, in: Abdul-Muhmin, 2003). Relationship satisfaction is conceptually different from product- or transaction-specific satisfaction. Seen from a client's perspective, product- and transaction-specific satisfactions deal with a client's experiences with specific episodic product or transaction encounters with an exchange partner, while relationship satisfaction has to do with the client's experience with the sum-total of product or service and transaction encounters over the lifespan of the relationship. Recall the IOR definition of Oliver (1990) and its emphasis on enduring linkages.

Although individual episodic encounters within this sum-total may have provided positive, negative, or neutral disconfirmations of expectations, it is the client's overall affect for the sum-total that is the focus of relationship satisfaction (Abdul-Muhmin, 2003: 620).

Geyskens et al. (1999) and Geyskens and Steenkamp (2000) distinguish two types of relationship satisfaction in marketing channel IORs: economic and social satisfaction. Both types make up "channel member satisfaction", which is defined as "a channel member's appraisal of all outcomes of its working relationship with another firm, including economic as well as social outcomes" (Geyskens and Steenkamp, 2000: 11). *Economic satisfaction* is defined as a channel member's evaluation of the economic outcomes that flow from the relationship with its partner such as sales volume, margins, and discounts. According to Geyskens et al. (1999: 224), an economically satisfied channel member considers the relationship to be a success in terms of realisation of economic or financial goals. It is satisfied with the general effectiveness and productivity of the relationship with its partner, as well as with the resulting financial outcomes. Researchers that have taken an economic view of satisfaction have defined it for example as a channel member's response to the perceived discrepancy between prior expectations and profits and the degree to which a firm's expectations towards the realisation of financial targets are met in the relationship.

*Social satisfaction* is defined by Geyskens and Steenkamp (2000: 13) as "a channel member's evaluation of the psychosocial aspects of its relationship, in that interactions with the exchange partner are fulfilling, gratifying, and facile". A channel member that is satisfied with the social outcomes of the relationship "appreciates the contacts with its partner, and, on a personal level, likes working with it, because it believes the partner is concerned, respectful, and willing to exchange ideas" (Geyskens et al., 1999: 224). Researchers that have considered satisfaction in more social terms have defined it as an evaluation of interaction experiences and the extent to which social interactions are gratifying.

A final definition of relationship satisfaction that we will discuss here is the frequently used one of Anderson and Narus (1990). Similar to Geyskens and Steenkamp (2000), Anderson and Narus have made a distinction between "rational" or economic aspects and social aspects of relationship satisfaction. First, they introduce the concept "outcomes given comparison level", which is defined as "a firm's assessment of the results (rewards obtained minus costs incurred) from a given working partnership in comparison with expectations based on present and past experience with similar relationships, and knowledge of other firms' relationships" (p. 44). Notice the similarity with the above stated economic view of satisfaction: the role of expectations on the one hand and profits, performance and outcomes on the other. Second, Anderson and Narus (1990: 45) define relationship satisfaction as "a positive affective state resulting from the appraisal of all aspects of a firm's relationship with another firm". Notice the parallel with Geyskens and Steenkamp's (2000) definition of channel member satisfaction. Anderson and Narus, however, emphasize the affective nature of relationship satisfaction *per se*, whereas Geyskens and Steenkamp view psychosocial aspects of a relationship as a separate dimension of relationship satisfaction.

*The relationship satisfaction construct in this study*

In this section we will evaluate the above definitions of relationship satisfaction, after which we will construct a new definition. This definition has been used as a lens with which the existing literature has been selected and reviewed.

If we examine the above definitions, we make five observations. First, the affective nature of satisfaction is mentioned several times. There is some debate on the importance of affect versus cognition in consumer decision making (e.g., Shiv and Fedorikhin, 1999) and in satisfaction (e.g., Oliver, 1993). Cognition seems to predominate and the role of affect seems to have received less attention. Through various studies – e.g., Oliver (1993) –, however, this imbalance has been redressed. Consequently, we regard satisfaction as an affective state (Anderson and Narus, 1990: 45; Masthoff and Gatt, 2006: 283), which is affected by cognitive processes.

Second, the definitions differ in the scope of the aspects that are taken into account when determining the satisfaction with a relationship. For instance, Anderson and Narus (1990) address *all aspects of a firm's relationship*, whereas Geyskens and Steenkamp (1999) focus on the economic and social dimensions. Because it may be too ambitious to assess *all* aspects of a relationship, and because ITO involves economic as well as social exchanges, we limit our definition to the economic and social aspects of relationship satisfaction. However, we include a third dimension, which we coin "technical satisfaction". This dimension focuses on the delivery of the IT services by the vendor (cf. "satisfaction with the core service", Crosby and Stevens, 1987), but encompasses "technical aspects" on the client's side as well, such as the obligation to clearly specify the requirements for the outsourced IT services (cf. Koh et al., 2004: 363). With the inclusion of this dimension we are able to explain better inconsistencies such as vendors that meet required service levels but whose clients are still dissatisfied (cf. Kern, 1997: 48). Meeting service levels may lead to "technically satisfied" clients, but does not necessarily implicate "socially satisfied" clients, because these clients may perceive their vendors to be bad communicators, untrustworthy, etc.

Third, the definitions may be divided between those that view satisfaction as an evaluation process and those that view it as a static state. We take on a dynamic perspective towards relationship satisfaction. As explained, the benefits of such an approach are that it enables us to evaluate the reciprocal interactions between the participants in an ITO relationship, and the fluctuations in levels of satisfaction and dissent, over time. Thus, we are interested in ITO relationship satisfaction at different points in time.

Fourth, the definitions describe satisfaction in positive, negative as well as neutral terms. We follow the latter approach since we contend that it is relevant to assess the *level of* relationship satisfaction, because of the mentioned dynamic perspective. Thus, the level of relationship satisfaction may fluctuate over time and may be either positive or negative, indicating satisfaction and dissatisfaction respectively.

Finally, all definitions offer a dyadic perspective in that the satisfaction of both partners – in our case a client and vendor – is taken into account.

Based on the above discussion we define ITO relationship satisfaction as: *an affective state resulting from the appraisal of the economic, social and technical aspects of an existing ITO partner's relationship with the other partner at a given point in time.*

We have explained what a relational perspective is and defined the concept of ITO relationship satisfaction. With this background information we turn to the Method section, where we explain how the existing relevant ITO literature has been reviewed.

## Method

This paper intends to provide a state-of-the-art review of studies regarding satisfaction with ITO relationships. Furthermore, it intends to examine whether studies with a dyadic perspective on ITO relationship satisfaction lead to better or richer explanations for ITO success than studies that focus on the *client perspective* only. The latter means that we will leave out the few ITO papers with a *vendor only* perspective, since these articles do not examine the vendor's satisfaction with the ITO arrangement, but for example focus on how vendors' value propositions contribute to client satisfaction (Levina and Ross, 2003) and building trust in ITO relationships (Oza et al., 2006). See De Clercq and Rangarajan (2008) for an example of a paper concerned with vendor satisfaction, though not within the ITO field.

*Selection criteria*

We undertook an extensive literature review to identify recent empirical work done on ITO relationships and the use of satisfaction at the overall relationship level as (proxy for) the dependent variable in these studies. The studies had to meet the following five criteria.

First, we are only interested in IT outsourcing. IT is increasingly critical for the survival of organisations. This requires high performing service providers in terms of reliability and availability of IT services. IT is present throughout most of today's firms and interrelates with practically all their business processes. Although it varies with the type of IT services delivered – systems operations, applications development, etc. – clients may therefore become too dependent on their vendor (cf. Lee and Kim, 1999: 52). IT develops at a dizzying pace, making it difficult to keep access to the latest IT knowledge. Finally, large switching costs are associated with ITO decisions. These characteristics distinguish ITO from other forms of outsourcing such as cleaning, catering or accounting (cf. Dibbern et al., 2004: 9; Kern and Willcocks, 2002: 3; Delen, 2005: 33), making it risky to generalise across these outsourcing types.

Second, we limited the search period to 1988 until now because many ITO scholars acknowledge that the Kodak-IBM deal in 1989 more or less started the ITO industry (e.g., Dibbern et al., 2004: 9; McLaughlin and Peppard, 2006) and because the study of ITO relationships is still relatively new.

Third, we only considered articles that have used relationship satisfaction, equivalent to our definition or in a closely related manner such as *overall satisfaction*, as (one of the) dependent variable(s). This means we left out articles concerned with user satisfaction (e.g., Sengupta and Zviran, 1997) or other satisfaction measures that do not capture the organisation as level of analysis (e.g., Goo et al., 2008; Kim et al., 2003). User satisfaction is important to assess the effectiveness, reliability, etc. (see Bailey and Pearson, 1983) of specific IS functions, such as computer applications. However, measurements of user (information) satisfaction are less suitable for evaluating overall outsourcing success, because those methods are not designed to directly link overall satisfaction with vendors to end users. Rather, these measurements have information systems as their unit of analysis (cf. Yoon and Im, 2005: 60; Dibbern et al., 2004: 73).

Fourth, the studies needed to have at least one relationship variable – e.g., trust, commitment – as independent variable. This is a consequence of the relational perspective that is applied in this study. We argue that the organisational features of ITO partners – for example firm size and organisations' capabilities – are necessary but not sufficient to explain total variance in ITO success. Let us consider vendor capabilities. From the resource-based view (e.g., Wernerfelt, 1984) may be derived that clients are seeking for specific capabilities with a vendor; capabilities they do not have themselves. These capabilities may be present with a certain vendor, yet may only become relevant in case a (client-vendor) relationship exists. Consequently, articles that were limited to organisational features as independent variables (e.g., Poppo and Zenger, 1998; Lee et al., 2009) were left out.

Finally, we also omitted theoretical and/or conceptual articles – that did not involve any kind of empirical data collection and analysis – since we were interested in empirical studies only.

The results were analysed for an indication of relevance to the field of ITO. Because many outsourcing studies exist in a variety of fields, we were often able to quickly discard those articles falling outside of ITO studies. For example, articles dealing with business process outsourcing (BPO), human resources outsourcing (HRO), or logistics. After this initial analysis, we culled through the abstracts and the method and conclusion sections of the articles that remained. Often, these sections provided us with sufficient information about the dependent variable, methods and unit of analysis. Based on a quick reading of the articles, we were able to eliminate conceptual articles or articles that still focused on user satisfaction or other dependent variables, such as risks, relationship quality, governance, participants' capabilities, etc. Articles concerned with IS in general or IT projects rather than IT outsourcing – e.g., Sabherwal (1999) and Taylor (2007) – were rejected as well.

When multiple articles appeared to have the same or similar authors, our criterion for selection was that the research had to be conducted on a unique database that was not repeated in other articles. When this occurred, we selected the one article from that database that appeared to best match our search criteria. This means that we have included Kern (1997) in favour of Kern and Willcocks (2000, 2002) and that we have included four articles of which Jae-Nam Lee is the (co-)author. The exception to this rule has been the work of Goles (2001, 2003). In his paper from 2003, he uses the client set of data of his research from 2001. We have included both studies so as to highlight the differences between a client only perspective and a dyadic perspective.

*Literature selection*

We first conducted a search for studies utilising both ABI/Inform Global (http://proquest.umi.com) and Web of Science (http://apps.isiknowledge.com). These are commonly used databases (cf. Lacity et al., 2009; Webster and Watson, 2002; Schwartz and Russo, 2004; Gonzalez et al., 2006; Oerlemans et al., 2007). Our search keys included *Information Technology outsourcing, Information Systems outsourcing, outsourcing,* and *satisfaction*. The results are shown in table 3 (Appendix A).

We carried out a second search by means of reviewing relevant IS related journals, as indicated by the literature review of Gonzalez et al. (2006), in order to cross-check the results of our first search. We applied the key word combinations from the first search in the search engines of the journals' websites. These key word combinations did not yield results with three journals. To make sure these journals really did not contain relevant articles we used broader key words as indicated in table 4. This search resulted in a small number of papers (e.g., Goo et al., 2008), not included in our earlier search. However, after glancing through the content of these papers, we discarded these articles since they did not meet the above-mentioned criteria.

Finally, we used earlier literature reviews – specifically Dibbern et al. (2004) and Sargent (2006) – and citations of already found articles in order to conclude our search for relevant studies.

*Index criteria*

Each article that met the above-mentioned criteria has been indexed. A summary has been produced for each study, which lists the year, country and sector in which the research has been undertaken, the type of empirical research conducted, the unit of analysis, the type of data (e.g., cross-sectional or longitudinal), the dependent variable / satisfaction construct used, and the key findings. The summaries provide us with easily identifiable indicators for comparisons of the reviewed studies.

First, through the *unit of analysis* used in each study we tried to identify the chosen viewpoint on ITO relationships, thus enabling us to distinguish between client – and dyadic perspective studies.

If the studies did not specify the unit of analysis, we turned to the type of *data collection* and its *content* as an indicator for the chosen perspective. The former includes whether data has been collected from client organisations only or from vendors as well and whether the data covers a certain period of time, or refers to a static moment in the relationship – recall that our definition of a relational perspective includes the time dimension. The latter includes the investigated types of outsourced IS functions – e.g., application development, data centres, end-user support. ITO literature (e.g., Kern, 1997: 53; Koh et al., 2004: 372) indicates that ITO relationships are dependent on the complexity of what is actually outsourced. The distinction between the various types may be an important differentiator between the two perspectives. E.g., one could argue that a dyadic perspective becomes even more relevant in case the type of outsourcing requires more interaction between the client and vendor and/or in case it is of higher value to the client organisation.

Third, we included the country and sector under research, since they may be a differentiator between the two perspectives. This is maybe relevant because it may be discovered that research in one country focuses more on the understanding of an ITO relationship in terms of a dyadic perspective than research in another country, which may be due to cultural differences. Furthermore, cultural differences may lead to differences in emphasis on specific relational variables, such as trust. The inclusion of demographic data also gives an indication as to what countries and sectors (public versus private sector, industry type, etc.) have been researched as far as ITO relationships are concerned.

Fourth, the type of empirical research (analytic, descriptive, etc.) is relevant because it shows in what way the studies try to capture the functioning of ITO relationships. This gives us an indication of the progress the research stream has made until now. For example, can signs of converging thought be discovered? Studies have been coded "analytic" in case they have been designed to examine associations between independent and dependent variables through a causal model and/or hypothesis testing. For example, causal models used in analytic studies show the constructs that are applied. These may be different across the perspectives, which in turn may lead to differences in explanations for ITO success.

Finally, the dependent variables (or satisfaction constructs) applied in the studies show the various definitions of the success of an ITO relationship. These definitions may differ across the perspectives, providing an additional insight in how the two types of studies examine ITO relationships and their outcomes.

*Comparison process*

With the above index criteria we were able to divide the selected studies between the client – and dyadic perspective groups. Next, we compared the two groups of studies on several aspects, including the index criteria. In order to assess whether the dyadic studies produce better or richer explanations for ITO success than the client-oriented studies, we needed to develop evaluation criteria with which to carry out this assessment. We have applied five evaluation criteria.

In addition to the dependent and independent variables, used in the reviewed literature, three criteria were taken from our discussion regarding the relational perspective. These three criteria are: longitudinal scope, dissent, and mutuality. *Longitudinal scope* was taken for its ability to assess fluctuations in relationship satisfaction over time. *Dissent* was chosen as it allows for dyadic partners to disagree on the level of satisfaction with a relationship. The *mutuality* concept acknowledges that ITO relationships constitute reciprocal processes between the dyadic partners (cf. Lacity et al, 2009: 137-9).

Table 1 gives a description of the criteria and, if applicable, their source. The results of the comparison are stated in the next section.

**Table 1    Measures with which to compare client – and dyadic perspective studies**

| Criteria | Description | Source |
| --- | --- | --- |
| Longitudinal scope | The time dimension is inherent in an ITO relationship. It allows for:<br>- fluctuations in level of relationship satisfaction; and<br>- possible mutuality effects between clients and vendors. | Oliver (1990); Goles and Chin (2005) |
| Dissent | The extent to which clients and vendors (dis)agree on the level of satisfaction with the ITO relationship. | Adapted from Meeus and Oerlemans (2002) |
| Mutuality | An ITO relationship is an ongoing reciprocal process in which actions of an ITO partner are contingent on rewarding reactions from the other partner. | Adapted from social exchange theory (e.g., Blau, 1964 in: Das and Teng, 2002) |
| Dependent variable | Gives the measures used to construct ITO success. These measures may differ across the perspectives. | --- |
| Independent variables | Gives the measures used to explain variance in ITO success. These measures may differ across the perspectives. | --- |

## Results

A summary of our literature review findings is presented in tables 5a and 5b (see Appendix B). Based on the index criteria, these tables outline the characteristics of each of the 22 studies we included and summarise the key findings. Several distinct themes emerged in the empirical research we explored. First, *the major finding of this study* is that the vast majority of studies (16 out of 22) uses a client's perspective, thus ignoring the vendor's side of an ITO relationship. This may be concluded from the unit of analysis used in the reviewed studies and/or from the type of data collection. We will elaborate this finding in detail below.

Second, a reasonable cross-cultural representation can be observed. Please note the number of studies included from each country between brackets. In addition to the "usual suspects", the United States (7) and the United Kingdom (2), research from Australia (1), China (1), Japan (1), Korea (6), Singapore (1), Taiwan (1), Malaysia (1), and Spain (1) is included in this overview. This is quite uncommon to mainstream ITO literature, which mainly revolves around outsourcing arrangements from the US and, to a lesser extent, the UK (e.g., Dibbern et al., 2004: 90; Gonzalez et al., 2006: 828). We also noticed the near absence of studies from mainland Europe in this review, except for the study from Spain. Continental Europe – e.g., Germany, France, the Benelux countries, the Nordics – is an important economic region for the global ITO market. Therefore, one might expect plentiful studies from this region. This finding, however, is consistent with earlier research. For example, Gonzalez et al. (2006) found that ITO articles from European countries, other than the UK, are scarce.

With regard to the distribution of studies between the two perspectives the following observations are worth mentioning. Both perspectives include studies from the US, the UK, Korea, and other Asian countries. This may indicate that a dyadic perspective is not country-specific. If we take a closer look at the only two countries with a fairly significant number of studies, the US and Korea, we find an interesting pattern. Five out of the seven studies from the US and five out of the six from Korea are from the client's perspective. One could thus assume that North-American and Korean researchers have a cultural preference to study ITO relationships from the client's point of view. This corresponds maybe with the individualistic nature of US culture. It does not, however, correspond with the collectivistic nature of an Asian country such as Korea (Hofstede, 2002; in Kor and Wijnen, 2005). It may therefore be true that the dyadic perspective is just a new research angle that will be increasingly applied in the near future (cf. Gonzalez et al., 2006: 827). The findings on the publication dates are an indication for this argument. We used a time frame of 22 years (1988-2010). As might have been expected, considering the novelty of the subject, the vast majority (17 out of 22) is from the last twelve years. More than half of the studies is from the last seven years; for the dyadic studies this is 50%. To summarise, given the preliminary nature of these findings, it is hard to draw decisive conclusions about the relationship between the country in which the research has been conducted and the chosen perspective.

Does this indecisiveness also apply to the sector under research? We think it does. Most studies (14) focus on the private sector, with just three articles solely concerned with the public sector (Harris et al., 1998; Lee, 2001; Willcocks and Kern, 1998) and five studies focused on both public and private sector. If we look at the dyadic perspective exclusively, it may be noticed that all studies are concerned with the private sector, with the exception of Koh et al. (2004) who have also included government agencies in their study. This means that the dyadic studies have an even greater bias towards the private sector than the studies on ITO relationships in general. We are not sure why this is the case. It may be true that ITO studies in general mainly concern with private sector organisations (cf. Hancox and Hackney, 2000: 217).

Fourth, the majority of the reviewed studies apparently do not deem the type of outsourced IS function to be important for studying ITO relationships. There is no clear difference between the unilateral and dyadic studies in this respect. Over one third of the studies (8) has not considered this topic at all and eight studies mention (almost) the whole range of IS functions without making inferences. The remaining studies, however, have taken the topic more seriously. Two articles (Willcocks and Kern, 1998; Marcolin and McLellan, 1998) have specifically focused on complex ITO relationships, something we would have expected more, assuming that the more complex an ITO relationship the more important that relationship becomes to the success of outsourcing. Two other articles (Susarla et al., 2003; Lee et al., 2007) have researched a specific IS function: Application Service Provision (ASP). To finish, two articles have examined the various IS functions in-depth. Grover et al. (1996) have found that telecommunications management and systems operations are positively related to ITO success, which was not the case for other IS functions. Seddon et al. (2007) did not link IS functions with ITO success, but have produced an extensive overview of characteristics associated with the IS functions – "considered for outsourcing", percentage outsourced, etc.

To conclude, we analysed the methodological aspects of the studies in more detail. Although all reviewed studies are analytic in nature, by which we mean they all consist of a research model to some extent, the studies differ considerably in terms of the used research models, the independent variables, and their operationalisations. This aspect does not differ across the perspectives. It makes it somewhat difficult to interpret and aggregate the findings from the reviewed studies (cf. Kern and Willcocks, 2002: 4; Dibbern et al., 2004: 67; Goles and Chin, 2005: 50). To give an example of the differences in research approach, consider the broad range of the reviewed studies. The studies range from the impact of contract characteristics on satisfaction (e.g., Harris et al., 1998; Marcolin and McLellan, 1998), via the analysis of the validity of existing models in an ITO context (e.g., Seddon et al., 2007; Yoon and Im, 2005; Koh et al., 2004; Susarla et al., 2003; Willcocks and Kern, 1998) to the influence of partnership variables on outsourcing success (e.g., Lee and Kim, 1999 & 2005; Grover et al., 1996).

In addition to the above comparison, we compared the two groups of studies through the five evaluation criteria as indicated in table 1. Here we will discuss our findings based on the comparison on these five criteria.

*Longitudinal scope*

The reviewed studies include only two longitudinal studies, one from the client perspective (Willcocks and Kern, 1998) and one from the dyadic perspective (Marcolin and McLellan, 1998). This is striking because of the continuous calls for longitudinal ITO research (e.g., Kern, 1997; Lee and Kim, 1999, 2005; Goles, 2001; Lee et al., 2007; Koh et al., 2004; Dibbern et al., 2004; Susarla et al., 2003). In terms of the division between the two perspectives, we may conclude that the dyadic perspective does not really "perform" better than the client perspective, although the frequency is higher (1 in 6 versus 1 in 16). As we would have expected both studies present results that indicate fluctuations in relationship satisfaction over time and the factors influencing it.

*Dissent*

*Satisfaction constructs used and dissent*

The vast majority of the reviewed studies measure satisfaction solely from the client's perspective. Table 5a indicates that they implicitly or explicitly consider client satisfaction to be equal to ITO success and (therefore) examine ITO arrangements through the client's perspective. The majority of these studies do not substantiate this choice made. Some authors only explain why satisfaction is an adequate measure for ITO success (e.g., Goles, 2003; Grover et al., 1996; Lee, 2001; Susarla et al., 2003) and neglect the possibility that there may be two sides to the satisfaction coin. Others (e.g., Hamaya, 2006; Harris et al., 1998) simply do not give an explanation why they have used client satisfaction as a proxy for the dependent variable ITO success. Consequently, these studies ignore the possibility that clients and vendors may perceive the level of satisfaction with the relationship differently – called *dissent*, which in turn may influence that very level of satisfaction in the ongoing relationship.

*Previous research on dissent*

The inaccuracy of a one-sided approach has already been debated in the marketing literature in the early 1980's. Before that period, studies of marketing channel dyads relied on measures obtained from one firm in a channel dyad. According to John and Reve (1982: 517), the common criticism at that time was that these measures were unable to produce valid tests of dyadic relationships. Hence, John and Reve researched cross-sectional channel dyads – wholesalers and retailers – from a *joint* perspective (Anderson and Narus, 1990: 43). Their research goal was to assess the construct validity of such an approach. John and Reve found that their measures for assessing the structure of a dyad showed convergent and discriminant validity, i.e. two types of construct validity. *Convergent validity* refers to the degree to which multiple attempts to measure the same concept by *maximally different* methods are in agreement (cf. John and Reve, 1982: 520; Bagozzi et al., 1991: 425). *Discriminant validity* is the degree to which measures of different concepts are distinct. The notion is that if two or more concepts are unique, then valid measures of each should not correlate too highly (Bagozzi et al., 1991: 425).

John and Reve also discovered that their measures for "dyadic sentiments", in terms of compatible goals, mutual trust, performance, satisfaction, and mutual agreement on roles, did *not* show adequate validity. The latter may not come as a surprise. John and Reve acknowledge this by concluding that informants from one side of a dyad may have different perceptions toward the relationship than those from the other side. Ye (2005) has in part used the same research method as John and Reve (1982) – taking a paired client-vendor sample to conduct path analysis, in addition to a client sample – and more or less comes to the same conclusion. Different perceptions may exist in an ITO context as well. Outsourcing clients and vendors may have divergent interests in the outsourcing arrangement (Kern, 1997: 51; Koh et al., 2004: 364; Wüllenweber et al., 2008: 4).

The incongruous outcome of John and Reve's study may be why Anderson and Narus (1990) have not replicated this "joint perspective" approach. Rather, they assessed both sides of the dyad separately, after which they determined commonalities and differences across both positions. Nonetheless, addressing both perspectives was the leading argument, since "both firms are participants in the same exchange relationship and … symmetry is expected in the *behavioural constructs* that underlie the relationship, [but] differences … are expected in terms of the indicators that reflect these constructs and the presence or relative strengths of the posited construct relationships" (p. 43). Anderson and Narus' findings corroborate this statement: out of the eighteen predicted relationships between ten constructs only six relationships were confirmed by both sides of the dyad. This is an important finding in itself, in that it clearly indicates the different perceptions on both sides of a dyadic relationship (i.e. dissent).

*Results on dissent in our study*

Review of the articles clearly shows the different results both perspectives produce with regards to dissent. Whereas the client perspective studies do not consider the concept of dissent to be pertinent, as may be concluded from the complete lack of discussion of the subject, the dyadic studies offer indications as to where differences of opinion between clients and vendors may arise. We will discuss a few examples here.

There are indications that vendors view an ITO relationship more as a partnership, whereas clients perceive it more as an arm's length arrangement (Goles, 2001: 135; Kern, 1997: 47). Although this finding may be ascribed to respondent selection bias – toward large private sector organisations – it is a theme consistent across the dyadic studies in this review, except Marcolin and McLellan (1998). It may seem positive that vendors perceive their relationships as partnerships. However, it might be argued (cf. Kern, 1997: 52; Lee et al., 2008: 158) that this is merely because vendors hope to increase their turnover with such a partnership. Clients, on the other hand, expect vendors to take charge in the delivery of services (Koh et al., 2004) and to manage the relationship, more or less "abdicating total responsibility" (Goles, 2001: 127). According to Ye (2005: 207), this may be an explanation for his finding that trust has a *negative effect* on ITO success, which is counterintuitive and inconsistent with other findings in ITO and IOR literature (see table 2). High levels of trust may reduce the client firm's willingness to learn because it can rely too much on the vendor to do everything. In addition, when the client trusts the vendor and leaves all the work to be done to the vendor, its limited participation and involvement may result in unexpected outcomes due to the lack of supervision or performance evaluation. This explanation is corroborated by the conceptual work of Ring and Van de Ven (1994) and Granovetter (1985): the emergence of trust on the part of a client is not sufficient to guarantee trustworthy behaviour by the vendor. There are also indications that vendors tend to have more positive perceptions of ITO relationships than clients, in terms of high ratings on trust, social interaction, and shared vision (Ye, 2005: 163-4). Ye suggests that vendors may be too overoptimistic about their relationships with the clients when the clients are actually less satisfied with the relationships. The degree of difference in perceptions, as indicated by Ye (2005), may pose a threat on ITO relationships.

*Independent variables*

The reviewed studies differ considerably in terms of the used research models. Thus, in order to compare the two groups of studies with regard to the independent variables, we need a common theme. Since all reviewed studies have in common that they include one or more relationship variables as independent variables, we focus on this aspect.

The majority of key findings in table 5a shows statistical evidence for the association between various relationship variables – e.g., trust, communication, commitment – and ITO success (in terms of client satisfaction). These findings imply that if one wants to have a successful ITO relationship, one has to address these relationship variables adequately. This correlation, however, has been measured by using the client's perspective only. Consequently, it may be interesting to determine what the results are from dyadic studies. Thus, we compare the findings from the studies in table 5b with those from table 5a. To this end, we selected a number of common relationship variables across the perspectives. The results are shown in table 2.

Although preliminary in nature, from table 2 we may learn that the findings from the unilateral studies are not as clear-cut as they may have seemed before this comparison. One would assume that for example higher levels of communication, cooperation, and knowledge sharing on both sides of a relationship would lead to higher satisfaction with that relationship. And because of the chosen research design, the unilateral studies imply that the findings are valid across an ITO relationship. However, the dyadic studies falsify these findings. These studies provide reasonable arguments why they have come to these counterintuitive results. We do not go into detail about these arguments, because we contend that these incongruent discoveries are interesting as such. That is, they provide evidence that clients and vendors perceive an ITO relationship differently in some respects, which may impact ITO success, as discussed with the concept of dissent. From the dyadic studies it seems that the authors involved hardly allow themselves to acknowledge this obvious but nonetheless important finding. Somewhat like the egg of Columbus, this obvious reality has apparently been overlooked by the vast majority of the reviewed studies.

**Table 2   Comparison of findings on (in)direct relationship variables from client – and dyadic perspective**

| Construct | Effect of the construct on ITO success, viewed from... | | |
|---|---|---|---|
| | Client perspective | Dyadic perspective | |
| | | Findings client | Findings vendor |
| Business understanding*[1] | Significant positive effect (Lee, 2001; Lee and Kim, 1999; Yoon and Im, 2005; Gonzalez et al., 2008) Not significant (Goles, 2003) | Not significant (Goles, 2001) | Significant positive effect (Goles 2001) |
| Commitment | Significant positive effect (Hamaya, 2005; Hussin et al., 2006; Lee, 2001; Lee and Kim, 1999, 2005; Sun et al., 2002) | Significant positive effect (Goles, 2001) | Significant positive effect (Goles, 2001) |
| Communication | Significant positive effect (Grover et al., 1996; Hamaya, 2005; Lee and Kim, 1999; Sun et al., 2002; Yoon and Im, 2005; Gonzalez et al., 2008) | Not significant (Goles, 2001) | Significant positive effect (Goles, 2001) |
| Cooperation | Significant positive effect (Grover et al., 1996; Lee and Kim, 2005) | Not significant (Goles, 2001) | Significant positive effect (Goles, 2001) |
| Coordination | Not significant (Lee and Kim, 1999) | Not significant (Goles, 2001) | Not significant (Goles 2001) |
| Cultural similarity | Not significant (Lee and Kim, 1999) | Not significant (Goles, 2001) | Not significant (Goles 2001) |
| (contract) Flexibility | Significant positive effect (Harris et al., 1998; Saunders et al., 1997*[2]) Not significant (Seddon et al., 2007) | Not significant (Goles, 2001) | Significant positive effect (Goles, 2001) |
| Interdependence | Significant negative effect (Lee and Kim, 1999, 2005) Not significant (Sun et al., 2002) | Significant negative effect (Goles, 2001; Lee et al., 2008) | Not significant (Goles, 2001; Lee et al., 2008) |
| Knowledge sharing | Significant positive effect (Hussin et al., 2006; Lee and Kim, 1999, 2005; Lee, 2001) | Significant positive effect (Koh et al., 2004; Lee et al., 2008) | Not significant (Koh et al., 2004) Significant positive effect (Lee et al., 2008) |
| Project monitoring by client | Significant positive effect (Hamaya, 2005) | n.a. | Significant positive effect (Koh et al., 2004) |
| Service quality | Significant positive effect (Goles, 2003; Grover et al., 1996; Hussin et al., 2006; Lee et al., 2007; Seddon et al., 2007; Susarla et al., 2003; Yoon & Im, 2005) | Significant positive effect (Goles, 2001) | Significant positive effect (Goles, 2001) |
| Top management support | Significant positive effect (Lee and Kim, 1999; Gonzalez et al., 2008) | n.a. | Significant positive effect (Koh et al., 2004) |
| Trust | Significant positive effect (Grover et al., 1996; Hussin et al., 2006; Lee, 2001; Lee and Kim, 1999, 2005; Sun et al., 2002`) | Significant positive effect (Goles, 2001; Lee et al., 2008*[3]) Significant negative effect *[4] (Ye, 2005) | Significant positive effect (Goles, 2001; Lee et al., 2008) Significant negative effect (Ye, 2005) |

*[1]  Lee and Kim (1999) view business understanding as a dimension of partnership quality, whereas Goles (2001, 2003) treats this construct as a vendor capability.
*[2]  We have interpreted Saunders et al.'s (1997) "tight contract" as being a "flexibel contract", in that it encompasses clauses that augment the degree of flexibility, such as clauses on growth rates, changes in business, etc.
*[3]  Lee et al. (2008) use the construct *mutual trust* as indirect variable for ITO success.
*[4]  This result was determined in the paired sample. At another point Ye (2005) indicates high ratings of vendors on trust (p. 164), making the conclusions somewhat contradictory.

*Inadequate treatment of relationship variables*

The above comparison of relationship variables brings a methodological issue to light. From the previous discussion we may learn that one has to be cautious to study inherently relational (cf. Zaheer et al., 1998: 143) – in our case dyadic – concepts, such as trust and communication, from one side of a dyad. And one has to be cautious with the interpretation of the findings. We explained in section 2 that these variables only exist when a relationship exists and that they entail reciprocal effects between dyadic partners, which may have effects on the outcomes of a relationship. Hence, in order to determine the impact of such relationship variables on the success of an ITO relationship we argued the necessity of assessing both clients' and vendors' opinions. After all, "it takes two to tango".

In contrast, all client perspective studies have treated relationship variables in a unilateral manner. We will give only one example, but the argument holds for all unilateral studies.

As part of their research models, Lee and Kim (1999, 2005) have examined the indirect impact of mutual dependency on ITO success. Contrary to their proposition, they found a negative link between mutual dependency and partnership quality – comprising of commitment, trust, etc. – in their 1999 study and again in 2005. Lee and Kim (1999) argued that the counterintuitive finding might be due to clients who may feel they become too dependent on a vendor after a certain period of time in an ITO relationship. In 2005 they ascribed it to the specific Korean circumstances of outsourcing IS functions within conglomerate groups, thus creating a "truck system" situation. However, if we take a closer look, the finding may not come as a surprise if one realizes that Lee and Kim did not measure mutual dependency, but *the dependency of the client*. And it may be true that clients feel they become too dependent on a vendor after a while. This does not, however, constitute a conclusive finding on the link between mutual dependency and ITO success. To that end, it would be advisable to use Anderson and Narus' (1990) method, as explained in section 2. In that case Lee and Kim might have come to the same conclusion as Goles (2001), in that clients may feel too dependent on the vendor, but the vendor may not feel the same, thus giving an incongruous outcome on the mutual dependency construct. In this case, a dyadic perspective does not lead to exceedingly different conclusions – e.g., a reverse finding – but produces more *depth* or contrast to the findings of a unilateral study.

*"Cross-boundary fallacy"*

This brings us to another methodological issue. It is also linked with the way in which the relationship variables have been treated, yet it is a more fundamental problem.

The vast majority of, if not all, reviewed studies have used key informants as an important source for research data. According to Bagozzi et al. (1991: 423), the key-informant method is a technique for collecting information on organisations and collectivities. Informants are chosen on the basis of particular qualifications such as specialised knowledge or position in an organisation. Rather than reporting on their own personal feelings or opinions, key informants provide information on the properties of organisations, their relationships with other organisations, etc. Bagozzi et al. report about the concerns researchers have expressed in recent years over the potential sources of measurement error in key informant responses. That is, key informants are often asked to perform complex judgment tasks dealing with organisational concepts such as environmental constraints, internal structure, power, and conflict. The question is whether key informants are able to perform these complex judgments without bias (cf. Ye, 2005: 165). We argue that this is specifically the case when key informants from client organisations are asked to make inferences about their vendors' perceptions, behaviours, etc. – e.g., Goles (2003); Lee and Kim (1999, 2005); Lee (2001). It seems reasonable to propose that if one wants to examine the "bigger picture" of an ITO relationship and/or the vendor perspective, a researcher should not rely on key informants from client organisations alone, but should also include key informants from vendor organisations.

*Working Paper – Do Not Cite or Quote without Author's Written Permission*

To summarise the above discussion, we introduce a new concept here. Analogous to Rousseau et al.'s (1998 in: Mouzas et al., 2007) *cross-level fallacy*, i.e. the inherent error to attribute individual motivations and behaviours to organisations, we call the problem discussed above *cross-boundary fallacy*. We define this concept as instances in which researchers mistakenly aggregate findings from one side of a dyad to the overall level of the dyadic relationship. This is merely another type of an external validity problem or the generalisability of the findings, of which construct validity is a sub type. Although not all client perspective studies have made this inaccuracy, one might have "suspicions" against them because of the use of the satisfaction construct at the overall relationship level.

*The dependent variable ITO success and its operationalisation*

From the review process it became clear that the operationalisation of the dependent variable ITO success needs further research (cf. Dibbern et al., 2004: 87). We will discuss this issue only briefly here by giving a few examples from the reviewed studies, as this topic is of such magnitude that it justifies a separate study.

Two indicators for ITO success are predominant in the reviewed studies. Grover et al.'s (1996) definition in terms of benefits attainment has been used seven times. Six studies make use of Anderson and Narus' (1990) satisfaction definition. Both indicators seem to be adequate with regard to the above-mentioned *cross-boundary fallacy*, in that with these measures informants are asked to reflect on their own perception of the relationship, not on the perceptions of the partner firm's staff. E.g., "our company's working relationship with Firm X has been an unhappy one" (Anderson and Narus, 1990: 51); "we are satisfied with our overall benefits from outsourcing" (Grover et al., 1996: 115). However, specifically with Anderson and Narus' construct it is questionable whether findings from using this measurement one-sidedly will produce valid results. It is of course relevant to know what one dyadic partner thinks of a relationship. But unless the sentiments of the other partner are asked for, it entails only half, or even less, of the story of that relationship. Recall that an ITO relationship constitutes: a) the client; b) the vendor; c) their exchange relationship; and d) the context. Since Grover et al. focus solely on the attainment of benefits by clients this flaw is circumvented. On the other hand, their measure involves other misconceptions.

Grover et al.'s (1996) study has been a pioneer one in the field of ITO relationships. They mention that the success dimension had not been operationalised until then. Thus, they acknowledge the preliminary nature of their measurements (p. 110). That said, other scholars continue to use Grover et al.'s construct without any adjustments, even until very recently (e.g., Lee et al., 2008; Hussin et al., 2006). This is in spite of arguments against the unadjusted application of the construct. For example, Grover et al.'s instrument assumes that the targeted benefits – e.g., focus on core business, access to skilled personnel, economies of scale – are relevant to all clients. Yet, not all clients have these goals with an ITO arrangement (Dibbern et al., 2004: 87; Seddon et al., 2007: 243). From a more theoretical point of view, Dibbern et al. (2004: 73) argue that the dependent and independent variables have to be assessed at the same level of analysis. Grover et al., however, measure the dependent variable success at the organisational level, whereas the independent variables are measured at the functional IS level.

Lee et al. (2008) and Ye (2005) not only have used Grover et al.'s instrument to determine the clients' perceptions about the success of an ITO arrangement, they also used the same construct to assess the vendors' perceptions on success. This is a rather surprising method, to put it mildly. Although it is understandable that the authors wanted to create symmetry in the success construct across the dyad for comparison purposes, this measure ignores the fact that vendors may have different indicators for success than those of their client. For instance, vendors may be predominantly interested in the turnover and profit margins of an ITO arrangement (cf. Lacity et al., 2008) or may focus on prompt payment by clients, as mentioned before.

A related topic is whether an ITO relationship involves a zero-sum game or not. It is argued repeatedly (e.g., Lacity, 2002; Lacity and Willcocks, 2003; Lacity et al., 2008) that "every dollar out of a client's pocket typically goes into the supplier's pocket". While this may be true, it may also be argued that new technologies or more efficient IT services, delivered by a service provider, generate additional dollars that typically go into the client's pocket. If for example a new internet technology, developed by a service provider and implemented in the client organisation, attracts more customers the client's revenues increase. Therefore, we argue that clients and vendors should concentrate together on "baking a bigger pie rather than trying to get a bigger slice from a fixed pie", isolating the notion of a zero-sum game.

A better way forward may be to analyse and/or replicate the measurement for ITO success Goles (2001) has created. He has found that both clients and vendors consider ITO success in terms of satisfaction, benefits attained and equity and that all three conditions must be met for both parties.

To conclude, the point we would like to make here is that satisfaction can best be used as a close proxy for overall ITO success when it is used from a dyadic perspective and in the way we have defined it. In so doing, it touches upon the benefits attainment and equity concepts, set forth by Goles (2001). Still then the results on a satisfaction score need to be examined in-depth. It is interesting to determine whether ITO partners are (dis)satisfied with their relationship. It is however more interesting to find out the reasons behind their (dis)satisfaction. For example, a vendor may meet all service level requirements, but a client may still be dissatisfied. Other, possibly more "soft" aspects may be important as well (cf. Kern, 1997: 48, 53). Therefore, for a more in-depth analysis it may be beneficial to consider other measures, such as benefits attained and equity, as well. This makes ITO success a multi-faced and complex construct, which is probably why *the* ITO success construct – such as the DeLone-Mclean (1992) *IS success* construct – has not been created yet.

*Mutuality*

The concept of mutuality and/or reciprocity runs through the above discussion of the four other evaluation measures. For example, without an enduring relationship (i.e. the *longitudinal scope*) reciprocal processes would not take place. Second, mutuality is inherent in the dissent concept, otherwise difference of opinion would not be a problem. It is the mutuality between clients and vendors that will eventually lead to discussions about the differences of opinion. Furthermore, we have shown that studying relationship variables from a dyadic perspective leads to a more in-depth understanding of such complex variables as (mutual) trust, (mutual) dependency, the inherently two-sided communication construct, etc. Finally, we have explained the mutual character of the dependent variable ITO success. Therefore, we will not examine this concept any further, except for one example. It may additionally illustrate the importance of mutuality in ITO relationships.

Koh et al. (2004) have used the *psychological contract perspective* to study ITO relationships. A first principle of this perspective is the "recognition of mutuality of the parties involved in a contractual relationship" (p. 357), and the subsequent recognition of mutual (rather than one-sided) obligations. This offers a new perspective, showing not only vendors have obligations in an ITO relationship; clients have obligations to vendors as well. For instance, the obligation to clearly specify requirements for the IT services, to closely monitor project progress, etc. Unilateral articles, such as Hamaya (2006) and Harris et al. (1998), have also pointed at the fact that clients have to adapt to outsourcing, and cannot pass on the "IT management burden" to the vendor entirely. Yet, they do not consider this to be an obligation of the client to the vendor, rather as an obligation to the clients themselves. That is, a client's "commitment to the request for proposal" (Hamaya) and "IS department flexibility" (Harris et al.) lead to better outcomes for the client. Thus, the unilateral studies have not considered a client obligation such as "prompt payment", like Koh et al. have done. While from a supplier's point of view this may be an important driver to sustain a relationship with a client.

*Summary of findings*

A clear conclusion that can be drawn from the empirical literature on satisfaction with ITO relationship is that it consists of a modest number of studies. And from the available studies just a handful uses a dyadic perspective. The vast majority of the reviewed studies is concerned with the client's perspective only. Second, inherently relational variables, such as communication and trust, have merely been treated as unilateral variables by these studies. Finally, the dependent variable ITO success needs further research since there are indications that this construct has been used inaccurately as well. Because of the fundamental nature of these findings, and because the studies differ significantly in terms of the used research models, etc. this study has mainly focused on methodological issues.

In summary, to give a preliminary answer to the stated research question, we contend that the dyadic studies produce a significantly better view on how underlying mechanisms of ITO relationships work, the distinctive roles clients and vendors play, and their possible impact on overall ITO success. On the other hand, it may be too early to assess whether dyadic studies produce better explanations for ITO success than client perspective research. The comparison we made between the two types of studies is by no means intended to be exhaustive. The reviewed studies differ too much and the number of dyadic studies is too few for that purpose. It does, however, intend to show that a dyadic perspective enriches the findings gathered from the unilateral perspective, in that it adds more depth or contrast to these findings, just as looking with two eyes in stead of one eye does.

## Future Directions for ITO Relationship Satisfaction Research

This section provides some guidelines for future research. We start with recommendations on the methodological issues discussed before that must be overcome if significant progress is to be made on the topic of ITO relationship satisfaction.

With regard to the client perspective bias our recommendation is simple and clear: when conducting research on ITO success at the overall relationship level, researchers should consider including both client's and vendor's perspectives. This will probably enhance the validity of the findings from such research. The client perspective only provides half – or even less – of the story of an ITO relationship, just as a husband or wife will mainly tell his/her side of the story of their marriage. Following this recommendation, the probability will increase that relationship variables in these studies will be treated correctly. Furthermore, it would be very beneficial to ITO research if the current problems with measuring ITO success would be solved. As said, we recommend analysing and/or replicating Goles' (2001) ITO success construct.

Other areas where future researchers may produce valuable contributions are threefold. First, we would like to reemphasize the earlier calls made about the relevant contributions longitudinal ITO research may have. In this way fluctuations in satisfaction with the relationship may be captured in stead of a snapshot view at a certain point in time, mutuality effects may be taken into account, the causality between variables may be determined with a higher degree of certainty, etc.

Second, we agree with Dibbern et al. (2004) that the ITO literature is "maturing" as opposed to "being mature". We have found studies in the reviewed research stream that build on earlier work. We also found articles testing existing models from other research streams. However, we noticed a lack of consistent treatment of variables and research models. This is by and large due to the pioneer character of the reviewed research stream. And it might therefore not yet be the time to consolidate the existing literature. Still, we recommend "building a cumulative tradition" (Dibbern et al., 2004: 9 & 87).

Finally, the near absence of studies from continental Europe in this research stream calls for future research. Continental Europe is an important economic region for the global ITO market. Consequently, it is important to find out how ITO relationships are being judged in a European context. Furthermore, this might help to assess the generalisability of the findings from the reviewed research.

## Conclusion

This paper provides a comprehensive overview of studies regarding satisfaction with ITO relationships, reviewing empirical studies carried out during the past 22 years. Two groups of studies have been identified: a handful of "dyadic perspective" studies that examine ITO relationships from both the client's and vendor's point of view and a larger number of "client only perspective" research. The *relational perspective* has been used as theoretical lens. The contributions of this paper concentrate on methodological issues, but include an important analysis of the studies' findings and some practical considerations as well.

Several ITO scholars have pointed out *that* the vendors viewpoints should be included. This paper has made an effort to explain *why* this is. So, from a methodological standpoint this paper raises questions about the construct validity of the studies that have used a client only perspective and, as a result, about the validity and reliability of their empirical findings.

These concerns are substantiated by a comparison of the two groups of studies on the following dimensions: longitudinal scope, dissent, mutuality, use of relationship variables that affect ITO success, and the satisfaction constructs used. Although the findings from the comparison are of a preliminary nature, we discovered that a dyadic perspective produces better results than a client only perspective for dissent, mutuality, and use of relationship variables. It does so by giving more *depth* to the findings. For instance, whereas client perspective studies claim that communication, trust, etc. are important factors for overall ITO success, dyadic perspective studies show that clients and vendors may think differently about those factors. The dyadic perspective studies do not significantly produce better results for longitudinal scope and partly perform worse in terms of the satisfaction construct used. That is, some dyadic studies use client satisfaction as a proxy for overall satisfaction, thus erroneously including the vendor in this equation.

Therefore, we contend that a dyadic perspective offers a more adequate method for studying ITO relationships than a client only perspective, but we suggest that a relational perspective may be even better. In this way an ITO relationship may also be studied over time and would by definition include the overall satisfaction with the relationship on both sides as well.

Our review has mainly focused on validity and measurement issues relevant to ITO researchers. Yet, practitioners may also benefit from this study. For instance, there are indications that practitioners have the same tendency to mainly focus on the client. This study provides preliminary support for an equal interest in vendors. After all, "it takes two to tango". As a result, it may be useful for practitioners to view ITO success from both sides of the relationship and design contracts accordingly.

Despite all the efforts we have made to present a complete review of ITO relationship satisfaction research, the study has clear limitations. Admittedly, the search procedure we used was somewhat subjective, so that some studies that might be considered to focus on relationship satisfaction may not have been included. Our search process was narrowly focused but serves to point out the relative dearth of empirical literature on the topic, despite a considerable amount of discussion about the importance of ITO client-vendor relationships. Nevertheless, we are convinced that we have compiled the vast majority of the studies carried out on this subject. We have investigated this research in a way that should provide useful insights for future researchers choosing either to use a relational perspective, consolidating the existing literature in order to come to consistent research models or to focus on a further exploration of the ITO success construct.

# Appendix A

Table 3  Number of studies found in Web of Science and ABI/Inform, using specific key words

| Search key | Web of Science (topic) | Web of Science (title) | ABI Inform (subject)* | ABI Inform (document title)* |
|---|---|---|---|---|
| Outsourcing | 2,988 | 1,677 | 3,025 | 1,603 |
| Information Technology Outsourcing | 349 | 38 | 445 | 52 |
| Information Systems Outsourcing | 324 | 46 | 192 | 61 |
| Outsourcing AND satisfaction | 78 | 8 | 28 | 9 |
| Information Technology Outsourcing AND satisfaction | 20 | 0 | 7 | 0 |
| Information Systems Outsourcing AND satisfaction | 23 | 0 | 3 | 0 |

Note: January 1988 to May 2010. * Only journal papers considered.

Table 4  Number of studies found in refereed journals, using specific key words

| Journal title | Number selected | Search term | |
|---|---|---|---|
| | | Information Technology / IT outsourcing satisfaction | Information Systems / IS outsourcing satisfaction |
| Communications of the ACM* (as of 1994) | 0 | 103 / 112 | 110 / 114 |
| Decision Support Systems | 0 | 50 / 50 | 50 / 0 |
| European Journal of Information Systems | 2 | 48 / 48 | 48 / 48 |
| Industrial Management and Data Systems | 1 | 0 / 2 | 0 / 2 |
| Information & Management | 1 | 79 / 80 | 80 / 0 |
| Information Systems Research | 1 | 16 / 16 | 16 / 16 |
| International Journal of Information Management | 0 | 43 / 44 | 44 / 0 |
| Journal of Information Technology | 0 | 30 / 30 | 30 / 30 |
| Journal of Management Information Systems (JMIS) ** | 2 | 1 / 2 | 1 / 2 |
| Management Information Systems Quarterly*** (as of September 1994) | 1 | 0 | 0 |
| California Management Review (CMR) ** | 1 | 1 / 0 | 0 / 0 |
| Decision Sciences (DS) ** | 0 | 0 / 0 | 0 / 1 |
| Harvard Business Review | 0 | 0 / 6 | 1 / 4 |
| Long Range Planning | 0 | 32 / 34 | 31 / 0 |
| Management Science | 0 | 0 / 0 | 0 / 0 |
| Sloan Management Review/MIT Sloan Management Review | 0 | 38 / 43 | 36 / 35 |
| OMEGA | 0 | 23 / 26 | 25 / 0 |
| Organisation Science (as of 1990) | 0 | 0 / 1 | 0 / 1 |

Note: January 1988 to May 2010, unless mentioned otherwise. * Journal and proceedings only. ** Other key words. JMIS; CMR; DS: Information Technology / IT (out)sourcing; Information Systems / IS outsourcing. *** Manual search in title, no search engine available.

# Appendix B

**Table 5a  Summary of empirical ITO relationship satisfaction studies, January 1988 to May 2010 (<u>client perspective</u>)**

| Author(s) | Year | Country / sector | Type of Analysis | Unit of analysis | Data | Dependent variable / satisfaction construct | Key findings |
|---|---|---|---|---|---|---|---|
| Goles | 2003 | US – Various industries: Telecom, Energy, Manufacturing, Finance, Service, Publishing | Analytic – research on the relationship between vendor capabilities, IS quality and overall customer satisfaction | Not specified, but focus on client perspective | Cross-sectional – 175 completed survey questionnaires from client managers with frequent direct contact with vendors. Type of outsourced IS functions not specified. | Overall outsourcing success, in terms of customer's level of satisfaction. Defined as a positive affective state resulting from all aspects of the subject matter being evaluated (ref. Anderson and Narus, 1984, 1990). Measured by three different items concerning the overall customer's satisfaction with the ITO arrangement. | Customers evaluate a vendor primarily in terms of its technical capability, or its ability to provide IS functions in an efficient manner. Customers believe that a vendor's ability to manage the relationship between the firms plays a significant role in ITO success, but not as significant as technical capability. Customers do not believe that the vendor's understanding of the customer's business makes a significant contribution to IS quality. IS quality is an important variable, linking vendor capabilities with ITO success. |
| Gonzalez, Gasco, and Llopis | 2008 | Spain – Private sector | Analytic – research on 1) important factors to achieving IS outsourcing success; and 2) measures with which to assess outsourcing success | Not specified, but focus on client perspective | Cross-sectional – 329 completed survey questionnaires. Type of outsourced IS functions not specified. | IS outsourcing success, in terms of a client's overall satisfaction and its perceived benefits from outsourcing (economic, strategic, and technological). | Choosing the right provider ranks first as factor that positively affects outsourcing success. Business understanding comes in second place. Top management support and communication are less important, although they have a significantly positive effect. Satisfaction and perceived benefits are adequate indicators for outsourcing success. |
| Grover, Cheon and Teng | 1996 | US – Various industries such as Manufacturing, Banking/finance, Insurance, Health care, Utilities / energy, Retail / wholesale, Transportation | Analytic – research on the relationship between extent of outsourcing certain IT functions and ITO success, with mediating effects of service quality and partnership | Client organisation | Cross-sectional – 188 completed and valid survey questionnaires. Type of outsourced IS functions considered: 1) applications development & maintenance; 2) systems operations; 3) telecommunications management; 4) end-user support; 5) systems planning & management; 6) all IS functions. | Satisfaction with strategic, economic, technological and overall benefits from ITO. Measured by: refocus on core business; enhancement of IT competence; access to skilled personnel; economies of scale in human resources; economies of scale in technological resources; control of IS expenses; reduced risk of technological obsolescence; access to key IT technologies; satisfaction with overall benefits. | Both service quality and the establishment of elements of partnership, i.e. trust, communication, satisfaction, and cooperation, are important determinants of ITO success. A base relationship exists between outsourcing telecommunications management and systems operations and ITO success. This is not the case for outsourcing other functions (e.g., applications development). Results show that a positive effect of ITO only leads to satisfaction with the arrangement if accompanied by cultivating a partnership at the outset of the relationship. |
| Hamaya (only survey A considered) | 2005 | Japan – Private sector | Analytic – research on the collaborative nature of ITO between clients and vendors | Not specified, , but focus on client perspective | 126 completed survey questionnaires. Type of outsourced IS functions not specified. | Clients' level of satisfaction with system integrators (SI) | Level of satisfaction directly affected by: SI selection; communication; and involvement of client's management. Indirect factors are: the client's commitment to the request for proposal; client's capability to carry out information system planning; and the alignment of IT and strategy in the client's organisation. |

| Author(s) | Year | Country / sector | Type of Analysis | Unit of analysis | Data | Dependent variable / satisfaction construct | Key findings |
|---|---|---|---|---|---|---|---|
| Harris, Giunipero and Hult | 1998 | US – Public sector | Analytic – research on the relationship between organisational and contract flexibility and ITO satisfaction | Not specified, but focus on client perspective | 48 completed survey questionnaires. Type of outsourced IS functions not specified. | Outcome of ITO arrangements, via client's satisfaction with the outsourcing contract. Measured via supplier: response time; cost savings; service levels; technical competency, and innovation. | ITO contracts are written to provide flexibility needed to cope with the changing environment. Client management is highly satisfied with the results of flexible contract encounters. Public sector IS departments should allow more flexibility in their organisational structures. |
| Hussin, Ismail, Suhaimi and Karim | 2006 | Malaysia – Public and private sector | Analytic – research on the relationships between partnership quality, service quality and ITO arrangements, and ITO success | Clients at the organisational level | Cross-sectional – 143 completed and valid survey questionnaires from client IT managers. IS functions mentioned, not researched: appl. development; hardware maintenance; telecommunication /network. | The dependent variable, ITO success, is measured via Grover et al.'s (1996) satisfaction construct; *see above*. | 1) Of Grover et al.'s (1996) 3 success dimensions, "strategic benefits" merges with "economic benefits". 2) Strategic benefits and technological benefits are equally important. 3) Partnership quality is an important predictor for ITO success. Benefit and risk-sharing, trust, commitment and knowledge-sharing are the key areas. 4) Service quality is also important for ITO success. 5) Degree of outsourcing and contract duration are found not to be significant predictors for ITO success. |
| Lee, Lee, Kim and Lee | 2007 | Korea – Private sector (service industry; distribution; manufacturing; and electronics and communications) | Analytic – research on the relationship between Application Service Provider (ASP) utilization, in terms of ASP service features, and satisfaction and performance of small firms | ASP service | Cross-sectional – 273 completed survey questionnaires and 466 responses to an online survey from small firms (less than 50 employees). IS functions considered: applications development & maintenance via ASP services. | Overall satisfaction with ASP service, i.e. the extent of sufficiency based on the expectation level of clients. Measured by: 1) satisfaction with the current ASP services; 2) use intention of the current ASP services; 3) recommendation of the ASP to partner companies | Customer service and maintenance has the strongest relationship with satisfaction of ASP services. The ASP price is the next most significant determinant. The security and risk factor was not statistically important for customer satisfaction. Effective training programs have a positive influence on satisfaction and organisational performance. Satisfaction and effectiveness of training have significant impact on organisational performance. |
| Lee and Kim | 2005 | Korea – Private sector (distribution; manufacturing; banking / finance / insurance; research; construction; and transport / warehousing / communication) | Analytic – research testing three rival models to assess the relationship among the determinants of an ITO partnership and to identify the relationship between partnership-related variables and ITO success | ITO relationship between client and vendor, focusing on the customer's perception of the relationship | Cross-sectional – 1) Interviews with 7 IS experts to test face validity. 2) Pretest for internal validity, with 2 to 5 employees from 36 organisations. 3) 225 completed and valid surveys from IS execs. IS functions considered: 1) appl. development; 2) appl. maintenance; 3) data centre; 4) network; 5) desktop; 6) help desk; 7) IS consulting. | The dependent variable, ITO success, is measured in terms of business satisfaction – via Grover et al.'s (1996) satisfaction construct; *see above* – and user satisfaction. The latter is measured with an adapted version of the instruments used by Bailey and Pearson (1983) and Baroudi et al. (1986). These include: reliability of information, relevancy, timeliness, etc. | The model based on behavioural-attitudinal theory is more suitable for understanding ITO success than a simple 1-D model or a model based on the theory of reasoned action. Psychological variables (mutual benefits, commitment and trust) are important predictors for ITO success. Behavioural variables (knowledge sharing, mutual dependence and cooperation) also influence ITO success significantly, but through the intervening psychological variables. Results show a negative relationship between mutual dependence and the psychological variables. This is counterintuitive and inconsistent with previous research. The authors ascribe this result to Korea's unique ITO situation in which the IS companies of *chaebols* hold about 60% share of the ITO market. |

| Author(s) | Year | Country / sector | Type of Analysis | Unit of analysis | Data | Dependent variable / satisfaction construct | Key findings |
|---|---|---|---|---|---|---|---|
| Lee and Kim | 1999 | Korea – Private sector (distribution; manufacturing; banking / finance / insurance; research; construction; and transport / warehousing / communication) | Analytic – research on the relationship between partnership quality, its components and determinants, and ITO success | ITO relationship between client and vendor, focusing on the customer's perception of the relationship | Cross-sectional – Interviews with IS experts to confirm external and content validity. 148 interviews with respondents from 36 clients involved with 54 vendors.<br><br>Type of outsourced IS functions not specified. | ITO success is measured by business satisfaction – Grover et al.'s (1996) satisfaction construct – and user satisfaction. The latter is measured with an adapted version of the instruments used by Bailey and Pearson (1983) and Baroudi et al. (1986). These include: reliability of information, relevancy, timeliness, etc. | All partnership quality variables (trust, business understanding, benefit and risk share, commitment) except conflict were significantly related to ITO success. Age of relationship and mutual dependency had a negative effect on partnership quality. Non significant relationships with partnership quality were found for: joint action, coordination, cultural similarity. (Partial) support was found for the relationship between partnership quality and: participation, communication, information sharing, and top management support. |
| Lee | 2001 | Korea – Public sector | Analytic – research on the relationship between knowledge sharing and ITO success, with the mediating effects of partnership quality and organisational capability | ITO relationship between client and vendor, focusing on the customer's perception of the relationship | 1) Interviews with ITO experts for external validity. 2) 195 completed and valid surveys from IS managers of various government agencies<br><br>IS functions considered: 1) appl. development; 2) appl. maintenance; 3) data centre; 4) network; 5) desktop; 6) help desk; 7) IS consulting | The dependent variable, ITO success, is measured in terms of business satisfaction – via Grover et al.'s (1996) satisfaction construct; *see above*. | Knowledge sharing is significantly associated with ITO success. The ability of the service receiver to absorb the needed knowledge has a significant direct effect on ITO success. Partnership quality plays a critical role as a mediator between knowledge sharing and ITO success. |
| Saunders, Gebelt and Hu | 1997 | US – Public and private sector (Finance, Airlines, Utilities, Petroleum, Consumer goods manufacturing and Food processing) | Analytic – research on 3 determinants of ITO success: nature of the contract (tight - loose); perceptions towards the vendor (supplier - partner); and 3) the role of IS (commodity - core) | Client's perceptions towards the outsourcing arrangement | Cross-sectional – Interviews with client managers at 9 public and 25 private organisations. IS functions considered: 1) appl. development; 2) data centre; 3) telecom and network; 4) end user support; 5) technical support services. 69% on average outsourced. | ITO success, measured along 4 dimensions: 1) economic; 2) technological; 3) strategic; and 4) overall satisfaction. | Supplier type relationships are much more likely to be economically and strategically successful when a tight contract has been written for the agreement. Overall, partnership arrangements were more successful than pure supplier relationships, especially when combined with tight contracts.<br><br>ITO seems to be most successful when clients view IS as a core function rather than as a commodity. |
| Seddon, Cullen and Willcocks | 2007 | Australia – Public sector and various industries in the private sector | Analytic – research to provide a preliminary test of the validity of Domberger's theory of *The Contracting Organisation* for use in an ITO context | Not specified, but focus on client perspective | Cross-sectional – 235 completed survey questionnaire All types of IS functions considered. Infrastructure management most frequently outsourced. 28% on average of IS functions outsourced. | "Satisfaction of the purchasing organisation with IT Outsourcing", measured by two items: 1) overall satisfaction with the benefits from outsourcing 2) satisfaction with the performance of the service provider | Domberger's theory does seem valid in an ITO context. Two factors of Domberger's theory, "specialization" and "market discipline" (better service), were significantly associated with satisfaction. However, "flexibility" and "cost savings" were not important in explaining satisfaction with benefits from ITO. The latter is surprising, considering the emphasis on cost savings in earlier ITO literature. |

| Author(s) | Year | Country / sector | Type of Analysis | Unit of analysis | Data | Dependent variable / satisfaction construct | Key findings |
|---|---|---|---|---|---|---|---|
| Sun, Lin and Sun | 2002 | Taiwan – Public sector, Manufacturing and SMEs | Analytic – research based on social exchange theory to test a research model that investigates the factors influencing ITO partnerships | "Each organisation's IS outsourcing case" | Cross-sectional – 1) Interviews with CEO's or CIO's from 8 client organisations to validate the model. 2) 197 completed and valid survey questionnaires from client IS managers. Type of outsourced IS functions not specified. | The dependent variable of this research model is outsourcing partnership measured by the satisfaction of the outsourcing service receiver. The operational definition is adapted from Anderson and Narus (1984, 1990). | Trust has a significant positive effect on dependence. Outcomes given comparison level has a strong positive link with ITO satisfaction and trust. Outcomes given comparison level and trust have a significant negative effect on conflict. The link between trust and commitment is significant. Mutual understanding is positively associated with communication and trust. Dependence has no effect on commitment. Conflict shows a significant negative effect on ITO satisfaction. Commitment is positively associated with ITO satisfaction. Power is not significant. |
| Susarla, Barua and Whinston | 2003 | US – Private sector | Analytic – research to test a conceptual model of satisfaction with an ASP (Application Service Provider) | Not specified, but focus on client perspective | First phase: unstructured questionnaires to design the perceived provider performance construct. Second phase: 256 completed and valid survey questionnaires from decision makers in mid-market companies. IS function considered: appl. development & maintenance via ASP. | ITO success is evaluated via satisfaction with ASP. It is "a positive affective state resulting from the appraisal of all aspects of a firm's working relationship with another firm" (Anderson & Narus, 1984, 1990). Evaluated after the client has been using the ASP's services for a period of time. Single construct, capturing satisfaction of users with the working relationship with the ASP as well as satisfaction with the ASP's service. | Satisfaction with an ASP is negatively affected by disconfirmation (i.e. discrepancy between performance and expectation), and positively influenced by the perceived provider performance and one form of prior experience, namely "prior systems integration". Further, perceived provider performance is positively influenced by the functional capability of the ASP and the quality assurance by the ASP, but negatively influenced by the prior systems integration. Two other experience norms, "Prior Internet usage" and "maturity of internal IT" were not found to be significant predictors of either perceived provider performance or satisfaction. |
| Willcocks and Kern * | 1998 | UK – Public sector | Analytic/descriptive – Research to test two analytical frameworks: 1) decision-making; and 2) determinants of an ITO relationship. | ITO relationship from the client's perspective | Longitudinal – 1993 to 1997. Eight respondents from client side. Interviewed 3 to 4 times, from unstructured to structured format. All IS functions considered: total ITO | Satisfaction is "a positive affective state resulting from the appraisal of all aspects of a firm's working relationship with another firm" (Anderson and Narus, 1984, 1990). | No asymmetry of dependence in favour of the vendor was found. The client had key in-house capabilities for ensuring business requirements, making technology work, and supply management. In ITO partnerships contract management is 'necessary but not sufficient'. Relationship management is pivotal as well. The exploratory "relationship" framework has considerable applicability in terms of coverage of issues. |
| Yoon and Im | 2005 | Korea – Private sector | Analytic – research to develop and test an evaluation framework for ITO client satisfaction, based on Analytic Hierarchy Process (AHP). | Not specified, but focus on client perspective | 32 completed and valid survey questionnaires from 25 companies. IS functions considered: 1) ASP; 2) data centre; 3) systems operations; 4) appl. Development & maintenance; 5) hybrid. | ITO customer satisfaction, measured through 3 evaluation areas (consulting service satisfaction, costumer supporting service satisfaction, and performance satisfaction), each of which consist of 2 to 3 evaluation factors, 6 to 13 attributes and 16 to 19 measurements | Customers are satisfied with the consulting service quality (e.g., in terms of used methodology and risk management) and consulting human resource (e.g., in terms of business understanding and communication). They are dissatisfied with SLA quality (including realization of service levels) and education services from ITO vendors. Customers are most satisfied with application development & maintenance, followed by hosting & data centre operations. |

\* Koh et al. (2004: 357) contend that this article uses a dyadic perspective. However, Willcocks and Kern indicate that their data is collected solely from the client side (p. 33).

**Table 5b  Summary of empirical ITO relationship satisfaction studies, January 1988 to May 2010 (<u>dyadic perspective</u>)\***

| Author(s) | Year | Country / sector | Type of Analysis | Unit of analysis | Data | Dependent variable / satisfaction construct | Key findings |
|---|---|---|---|---|---|---|---|
| Goles | 2001 | US – Various industries: Telecom, Energy, Manufacturing, Finance, Service, Publishing | Analytic – research on the link between the vendor's and client's capabilities, their relationship, IS quality and overall satisfaction | ITO relationship between client and vendor, viewed from both the service receiver and provider's perspectives. | Cross-sectional – 175 completed survey questionnaires from client managers and 191 from vendor managers. Type of outsourced IS functions not specified. | Overall ITO success: satisfaction, benefits attained and equity, as perceived by both clients and vendors. Satisfaction refers to Anderson and Narus' (1990) definition. Benefits attained are evaluated via reflective measures as opposed to the formative measures of Grover et al. (1996). Equity refers to "a fair return for the efforts or resources provided". | This study shows significant differences in perceptions between clients and vendors with regard to the importance of the participants' capabilities and the relationship factors. Second, vendors and clients perceive their relationship differently. Vendors view it more as a partnership, whereas clients perceive it more as an arm's length arrangement. Finally, both customers and vendors consider ITO success in terms of satisfaction, benefits attained and equity. All three conditions must be met for both parties. |
| Kern | 1997 | UK – Private sector (6 firms in as many different industries) | Conceptual/analytic – research to test a theoretical model regarding the characteristics of an ITO relationship between client and vendor | ITO relationship between client and vendor | Cross-sectional – Interviews with 6 client (IT) managers and 6 vendor directors. 11 case studies conducted (document analysis). Type of outsourced IS functions mentioned, but not researched: 1) software development and operations; 2) legacy systems management; 3) data centre management; 4) all IS functions. | Satisfaction is "a positive affective state resulting from the appraisal of all aspects of a firm's working relationship with another firm" (Anderson and Narus, 1984, 1990). | The research model generally represents the main characteristics of an ITO relationship. It includes: exchanges (services, communication, etc.); factors such as shared vision and social bonds; a working context with elements like commitment and trust; and two relationship foci, i.e. contractual and normative. Large differences in clients' and vendors' perceptions exist. Vendors stronger emphasize the importance of partnering. Although SLA's are met by vendors, clients report dissatisfaction. Vendors need greater business understanding, show more commitment, and should invest beyond the agreed-upon terms. The study confirms that the relationship is a major, neglected subject in ITO literature. |
| Koh, Ang and Straub | 2004 | Singapore – Various industries such as Government, Banking, Retail, Health care, Transport and Manufacturing | Analytic – research to determine client-vendor obligations in ITO and survey questionnaires to assess the impact of fulfilling these obligations on ITO success | ITO relationship between client and vendor, viewed from the individual level of analysis: the perspectives of both clients' and vendors' project managers | Cross-sectional – 1) Interviews with 9 client project managers and 6 vendor managers. 2) 370 completed and valid surveys from 90 clients and 68 suppliers. IS functions mentioned: appl. development and maintenance | ITO success, operationalized through items for overall satisfaction with the contract as well as the desire to retain the outsourcing partner. Measurements used: satisfaction (adapted from Poppo and Lacity 2002) and intention to continue the outsourcing relationship (adapted from Kristensen et al. 2000). | Supplier obligations are: accurate project scoping, clear authority structures, taking charge, effective human capital management, effective knowledge transfer, and effective inter-organisational teams. Suppliers perceive customer obligations as clear specifications, prompt payment, close project monitoring, dedicated project staffing, knowledge sharing, and project ownership. ITO success shows significant relationship with fulfilling all of these obligations – except accurate project scoping, project staffing and knowledge sharing. |

| Author(s) | Year | Country / sector | Type of Analysis | Unit of analysis | Data | Dependent variable / satisfaction construct | Key findings |
|---|---|---|---|---|---|---|---|
| Lee, Huynh and Hirschheim | 2008 | Korea – Private sector. Various industries such as Banking and Finance, Manufacturing, Retail and Wholesale, Construction, Transport and Communication | Analytic – research on the relationship between various types of trust and knowledge sharing and ITO success | ITO relationship between client and vendor, viewed from both the service receiver and provider's perspectives. | Cross-sectional – 1) Interviews with 7 ITO experts to refine survey. 2) 163 completed and valid surveys from top IT execs in as many clients. 3) 45 completed and valid surveys from as many suppliers. Type of outsourced IS functions not specified. | The dependent variable, ITO success, is measured in terms of business satisfaction – via Grover et al.'s (1996) satisfaction construct; *see above*. | The results from this study partially support the proposed framework. Mutual trust between the client and vendor is very important for knowledge sharing and ITO success, and is affected by the initial perception to each other's partner at the beginning of the outsourcing process. Interestingly, initial trust is considered a significant factor in the perception of mutual trust from the client's perspective, but not from the vendor's viewpoint. |
| Marcolin and McLellan | 1998 | US – Banking | Analytic/descriptive – research to test Fitzgerald and Willcocks' (1995) model on the impact of uncertainty and contractual definition on type of relationship between client and vendor | ITO relationship between client and vendor | Longitudinal – Interviews with 27 bank managers and vendor managers. Type of outsourced IS functions: all types, with 75-100% outsourced. "Complex ITO". No results presented. | "Overall satisfaction with IS" as a measure of "better outsourcing arrangements". Satisfaction was measured by a 10-point Likert scale, with the extremes representing very satisfied and very dissatisfied outcomes. Satisfaction before and after outsourcing were measured. | Partnerships do not produce higher satisfaction. "Buyer/sellers" achieve greater satisfaction via more control and certainty. Strategic partners systematically look to build relationships, while buyer / sellers look to maximize their position. Overall, companies can successfully manage relationships in any situation, no matter how uncertain the environment, as long as they choose the right combination of business objectives, contractual stances and relationship management. |
| Ye | 2005 | China – Various industries such as Software and technology, Transportation and Logistics, Mining, Manufacturing, and Financial Services | Analytic – research on the link between three dimensions of social capital and ITO success, with the mediating role of knowledge acquisition | The focal firm embedded in a dyadic relationship | Cross-sectional – 1) Interviews with 4 client IT managers. 2) 230 completed and valid surveys from 151 client and 79 supplier organisations. IS functions researched (no results presented): 1) System Operations; 2) Telecom / Networks; 3) Appl Development & Maintenance; 4) User Support; 5) IS Planning & IS Management. | Two dimensions of ITO success: 1) success in business operations and 2) success in IT-enabled innovation. The first dimension is measured via a slightly adapted version of Grover et al.'s (1996) satisfaction construct**; *see above*. | Both social capital and knowledge acquisition are crucial to ITO success. Different aspects of social capital play different roles in the process of IT value creation. Specifically, the structural dimension (*partner resource endowment*) and the cognitive dimension (*shared vision* and *shared cognition*) have a strong impact on *knowledge acquisition*; whereas the relational dimension (*social interaction* and *trust*) has strong direct effects on successful outcomes of ITO. |

\* We came across another dyadic study (Oza, 2006) in our search process, but it did not use ITO success or its proxy satisfaction as dependent variable.
\*\* Ye (2005) contends that he uses items from Lee and Kim's (1999) instrument, which in fact are from Grover et al. (1996).